\documentclass[12pt]{article}
\usepackage{amsmath,amssymb}
\usepackage{graphicx} 
\usepackage{bbm}
\usepackage{longtable}
\usepackage[small,loose]{subfigure}
\usepackage{fleqn} 
\usepackage{mathrsfs}   
\usepackage[small,sf]{caption2}

\advance \headheight by 3.0truept       

\newcommand{\CenterObject}[1]{\ensuremath{\vcenter{\hbox{#1}}}}

\newcommand{\I}{\mathrm{i}}

\newcommand{\E}[1]{\ensuremath{\mathrm{E}_{#1}}} 

\newcommand{\SO}[1]{\ensuremath{\mathrm{SO}(#1)}}
\newcommand{\SU}[1]{\ensuremath{\mathrm{SU}(#1)}}

\newcommand{\Z}[1]{\ensuremath{\mathbbm{Z}_{#1}}} 

\allowdisplaybreaks[0]

\begin{document}
\title{
\hfill {\normalsize TUM-HEP-659/07}\\[2cm]
{\bf\Huge Mirage torsion}\\[0.8cm]}

\setcounter{footnote}{0}
\renewcommand{\thefootnote}{\alph{footnote}}

\author{{\bf\normalsize 
Felix~Pl\"oger$^{\boldsymbol{1},}$\footnote{\texttt{ploeger@th.physik.uni-bonn.de}},~
Sa\'ul~Ramos-S\'anchez$^{\boldsymbol{1},}$\footnote{\texttt{ramos@th.physik.uni-bonn.de}}},\\
{\bf\normalsize 
Michael Ratz$^{\boldsymbol{2},}$\footnote{\texttt{mratz@ph.tum.de}}~~and
Patrick~K.~S.~Vaudrevange$^{\boldsymbol{1},}$\footnote{\texttt{patrick@th.physik.uni-bonn.de}}}\\[1cm]
{\it\normalsize
${}^1$ Physikalisches Institut der Universit\"at Bonn, Nussallee 12, 53115 Bonn,
Germany.}\\[0.1cm]
{\it\normalsize
${}^2$ Physik Department T30, Technische Universit\"at M\"unchen, 85748 Garching,
Germany.}
}

\date{}

\maketitle \thispagestyle{empty} 

\abstract{%
$\Z{N}\times\Z{M}$ orbifold models admit the introduction of a discrete torsion
phase.  We find that models with discrete torsion have an alternative
description in terms of torsionless models. More specifically, discrete
torsion can be `gauged away' by changing the shifts by lattice vectors.
Similarly, a large class of the so-called generalized discrete torsion phases
can be traded for changing the background fields (Wilson lines) by lattice
vectors. We further observe that certain models with generalized discrete
torsion are equivalent to torsionless models with the same gauge embedding but
based on different compactification lattices. We also present a method of
classifying heterotic $\Z{N}\times\Z{M}$ orbifolds.
}

\setcounter{footnote}{0}
\renewcommand{\thefootnote}{\arabic{footnote}}

\clearpage

\section{Introduction}

Among the currently available frameworks, superstring theory appears to have the
greatest prospects for yielding a unified description of nature. Optimistically
one may hope to identify a string compactification that reproduces all
observations.
The perhaps simplest way to obtain a chiral spectrum in four dimensions, as
required by observation, is to compactify on an
orbifold~\cite{Dixon:1985jw,Dixon:1986jc}. 
Although it is straightforward to compute orbifold spectra, a deep understanding
of these constructions, including an interpretation of the zero-modes, is harder
to obtain. Obstructions arise from the large number of possible gauge embeddings
and geometries, as well as other degrees of freedom. The classification of gauge
embeddings has been accomplished only in prime orbifolds. The generalization to
$\Z{N}\times\Z{M}$ orbifolds with or without Wilson lines has not been discussed
in the literature so far. $\Z{N}\times\Z{M}$ orbifolds are particularly rich as
they can be generalized by turning on certain phases which are known as discrete
torsion~\cite{Vafa:1986wx,Font:1988mk,Vafa:1994rv,Sharpe:2000ki,Gaberdiel:2004vx}.

Aiming at a systematic understanding of heterotic $\Z{N}\times\Z{M}$ orbifolds, we set out
to survey the possibilities arising in these constructions. In the course of our
investigations we obtain rather surprising results. First of all, discrete
torsion can be `gauged away' in the sense that models with discrete torsion have
an alternative description in terms of torsionless models. Moreover, we shall
see that the so-called `non-factorizable' orbifolds are equivalent to
factorizable orbifolds with (generalized) discrete torsion or different gauge embedding. 

This paper is organized as follows. In section~\ref{sec:ZNxZM} we collect some
basic facts on the construction of $\Z{N}\times\Z{M}$ orbifold models. We
encourage readers who are familiar with the construction of orbifolds to
skip this section. In section~\ref{sec:DiscreteTorsion}, we establish the
equivalence between switching on a discrete torsion phase and changing the gauge
embedding by elements of the weight lattice.
Section~\ref{sec:GeneralizationDTAndBrothers} is devoted to the generalization
to orbifolds with Wilson lines. In section~\ref{sec:ClassifZ3xZ3} we outline a
prescription for a classification of $\Z{N}\times\Z{M}$ orbifold models.
Finally, section~\ref{sec:Conclusions} contains a discussion of our results.
Some issues concerning the transformation phases are discussed in the appendix.

\section{$\boldsymbol{\Z{N}\times\Z{M}}$ orbifold compactifications}
\label{sec:ZNxZM}

\subsection{Setup}

Let us start by reviewing some basic facts on orbifold compactifications
\cite{Dixon:1986jc}. To construct an orbifold, one first considers a
$d$-dimensional torus $\mathbbm{T}^d$, which can be understood as
$\mathbbm{R}^d/\Gamma$, i.e.\ as the $d$-dimensional space with points
differing by lattice vectors $e_\alpha \in \Gamma$ identified. In this study we will take
$d=6$ in order to arrive at an effective four-dimensional theory at low
energies. If the torus lattice enjoys one or more discrete rotational symmetries
comprising the point group $P$, one can define an orbifold as the quotient
$\mathbbm{O}=\mathbbm{T}^6/P$. Equivalently one can describe the orbifold by
\begin{equation}
 \mathbbm{O}~=~\frac{\mathbbm{R}^6}{S}\;,
\end{equation}
where $S$ is the space group.  Space group elements consist of discrete
rotations and translations by lattice vectors $e_\alpha$. We will be mostly
interested in $\Z{N}\times\Z{M}$ orbifolds, in which the torus lattice has two
discrete rotational symmetries described by the independent twists $\theta$ and
$\omega$, whereby $\theta^N=\omega^M=\mathbbm{1}$ and $N$ is a multiple of $M$.
Space group elements $g\in S$ are then given by  $g = (\theta^{k_1}\,\omega^{k_2},
n_\alpha\, e_\alpha)$ where $0\le k_1 \le N-1$, $0\le k_2\le M-1$ and
$n_\alpha\in\mathbbm{Z}$. 
Further, we restrict our analysis to models with \SU3 holonomy, where the
rotations can be diagonalized,
\begin{equation}
 \theta\,z^i~=~\exp(2\pi\I\,v_1^i)\,z^i
 \quad\text{and}\quad
 \omega\,z^i~=~\exp(2\pi\I\,v_2^i)\,z^i\;,
\end{equation}
with $z^{1,2,3}$ being the complex coordinates of the compact space, and $\sum_i
v_{1,2}^i=0$. Unless stated otherwise, we use 
\begin{equation}
 v_1~=~\frac{1}{N}(1,0,-1;0)
 \quad\text{and}\quad
 v_2~=~\frac{1}{M}(0,1,-1;0)\;.
\end{equation}

The space group action is to be embedded in the gauge degrees of freedom
according to
\begin{equation}
 g = (\theta^{k_1}\,\omega^{k_2},n_\alpha\,e_\alpha)~\hookrightarrow~
 (\mathbbm{1},V_g)\qquad V_g = k_1\,V_1+k_2\,V_2+n_\alpha\,A_\alpha\;,
\end{equation}
where $V_1$, $V_2$ are the shifts, $A_\alpha$ are the Wilson lines, and $V_g$ denotes the local shift corresponding to the twist $v_g = k_1 v_1 + k_2 v_2$. Due to the embedding, they have to be of appropriate orders:
\begin{equation}\label{eq:latticeconditions}
 N\, V_{1} ~\in~ \Lambda\;,\quad
 M\, V_{2} ~\in~ \Lambda\;,\quad
 N_{\alpha}\, A_{\alpha} ~\in~ \Lambda\;.
\end{equation}
Here $\Lambda$ is the $\E8\times\E8$ or $\mathrm{Spin}(32)/\Z2$ weight
lattice,\footnote{Since these lattices are self-dual, we denote the root and
weight lattice by the same symbol.}
and $N_\alpha$ denotes the order of the Wilson line $A_\alpha$, which is
constrained by geometry.

Modular invariance of one--loop amplitudes imposes strong conditions on the shifts and Wilson lines. In $\Z{N}$ orbifolds, the shift $V$ and the twist $v$ must fulfill~\cite{Dixon:1986jc,Vafa:1986wx}:
\begin{equation}\label{eq:znmodularinv}
N \,\left(V^2 - v^2\right)~=~0 \mod 2 \,.
\end{equation}
In $\Z{N}\times\Z{M}$ orbifolds with Wilson lines, modular invariance, together with consistency
requirements (see appendix \ref{app:ModularInvariance}), requires
\begin{subequations}\label{eq:newmodularinv}
\begin{eqnarray}
  N\,\left(V_{1}^{2} - v_{1}^{2} \right)  & = & 0 \mod 2\;, \\
  \label{eq:fsmi1}
  M\,\left(V_{2}^{2} - v_{2}^{2} \right)  & = & 0 \mod 2\;, \\
  \label{eq:fsmi2}
  M\,\left(V_{1}\cdot V_{2} - v_{1}\cdot v_{2} \right)  & = & 0 \mod 2\;, \\
  \label{eq:fsmi3}
  N_\alpha\,\left(A_{\alpha}\cdot V_{i}\right)  & = & 0 \mod 2\;, \\
  \label{eq:fsmi4}
  N_\alpha\,\left(A_{\alpha}^2\right)  & = & 0 \mod 2\;, \\
  \label{eq:fsmi5}
  Q_{\alpha\beta}\,\left(A_{\alpha}\cdot A_{\beta}\right)  & = & 0 \mod 2 \quad (\alpha \neq \beta)\;,
  \label{eq:fsmi6}
\end{eqnarray}
\end{subequations}
where $Q_{\alpha\beta}\equiv\text{gcd}(N_{\alpha},N_{\beta})$ denotes the greatest common divisor of
 $N_{\alpha}$ and $N_{\beta}$.

\subsection{Spectrum}

Given a compactification lattice, the discrete rotations described by $v_{1,2}$,
shifts and Wilson lines, there exists a standard procedure to calculate the
massless spectrum (cf.\
\cite{Ibanez:1987pj,Font:1989aj,Forste:2004ie,Kobayashi:2004ya,Buchmuller:2004hv,Buchmuller:2006ik}). The Hilbert space
decomposes in untwisted and various twisted sectors, denoted by $U$ and
$T_{(k_1,k_2,n_\alpha)}$, respectively. The gauge group after compactification
is generated by the 16 Cartan generators plus roots $p\in\Lambda$ ($p^2=2$)
fulfilling 
\begin{equation}
 p\cdot V_i~=~0\mod1 \quad\forall i\;,\quad 
 p\cdot A_\alpha~=~0\mod1\quad\forall\alpha\;.
\end{equation}
Chiral untwisted sector states are described by $p\in\Lambda$ and
$q\in\Lambda_{\SO8}$ ($q^2=1$) satisfying
\begin{subequations}
\begin{eqnarray}
 p\cdot V_i-q\cdot v_i & = &0\mod1\;,\quad i \in 1,2\;,\\
 q\cdot v_i & \ne & 0\;,\quad i=1\:\text{and/or}\:2\;, \\
 p\cdot A_\alpha & = & 0\mod 1\quad\forall\alpha\;.
\end{eqnarray}
\end{subequations}
Twisted sector zero modes are associated to the inequivalent `constructing
elements' $g=(\theta^{k_1}\omega^{k_2},n_\alpha e_\alpha)\in S$, corresponding
to the inequivalent fixed points and fixed planes. For each such $g$ one solves
the mass equations 
\begin{subequations}\label{eq:MassEquation}
\begin{eqnarray}
 \frac{1}{8}m_\mathrm{L}^2 
 & = & 
 \frac{1}{2}p_\mathrm{sh}^2-1+
 \omega_i\,\widetilde{N}_{g,i}
 +\overline{\omega}_i\,\widetilde{N}_{g,i}^*
 +\delta c ~\stackrel{!}{=}~0\;,\\
 \frac{1}{8}m_\mathrm{R}^2 
 & = &
 \frac{1}{2}q_\mathrm{sh}^2-\frac{1}{2}+\delta c
 ~\stackrel{!}{=}~0\;,
\end{eqnarray}
\end{subequations}
with the shifted momenta ($p\in\Lambda,q\in\Lambda_{\SO8}$)
\begin{subequations}
\begin{eqnarray}
 p_\mathrm{sh} & = & p+V_g\;,\\
 q_\mathrm{sh} & = & q+v_g\;.
\end{eqnarray}
\end{subequations} Here $\omega_i=(v_g)_i\mod1$  and
$\overline{\omega}_i=-(v_g)_i\mod1$,  such that
$0<\omega_i,\,\overline{\omega}_i\le1$. Moreover, $\widetilde{N}_{g,i}$ and
$\widetilde{N}_{g,i}^*$ are integer oscillator numbers. Finally, $\delta
c=\frac{1}{2}\sum_i\omega_i\left(1-\omega_i\right)$.
  
The states $|q_\mathrm{sh}\rangle_\mathrm{R} \otimes |p_\mathrm{sh}\rangle_\mathrm{L}$, where
$q_\mathrm{sh}$ and $p_\mathrm{sh}$ are solutions of the mass
equations~(\ref{eq:MassEquation}), are subject to certain invariance conditions: commuting elements
$h=(\theta^{t_1}\omega^{t_2},m_\alpha e_\alpha)$, with $[g,h]=0$, have to
act as the identity on physical states. This leads to the projection condition 
\begin{equation}\label{eq:transformation}
|q_\mathrm{sh}\rangle_\mathrm{R} \otimes |p_\mathrm{sh}\rangle_\mathrm{L}
~\stackrel{h}{\longmapsto}~
\Phi\, |q_\mathrm{sh}\rangle_\mathrm{R} \otimes |p_\mathrm{sh}\rangle_\mathrm{L}
~\stackrel{!}{=}~  
|q_\mathrm{sh}\rangle_\mathrm{R} \otimes |p_\mathrm{sh}\rangle_\mathrm{L}\;.
\end{equation}
Here the transformation phase $\Phi$ is given by 
\begin{equation}\label{eq:transformationphase}
 \Phi
 ~\equiv~
 e^{2\pi\I\,[p_\mathrm{sh}\cdot V_h- q_\mathrm{sh} \cdot v_h 
 + (\widetilde{N}_g - \widetilde{N}_g^*)\cdot v_h]}\,
 \Phi_\mathrm{vac}\;,
\end{equation}
where (cf.\ appendix \ref{app:ModularInvariance})
\begin{equation}
 \Phi_\mathrm{vac}
 ~=~
 e^{2\pi\I\,[-\frac{1}{2}(V_g\cdot V_h - v_g\cdot v_h)]}\;.
\end{equation}
Equation \eqref{eq:transformation} states that the transformation phase $\Phi$
has to vanish, which will be important for the following discussion.

\section{Brother models and discrete torsion}
\label{sec:DiscreteTorsion}

In this section we start by examining a new possibility to find inequivalent
models.  We discuss under what circumstances models with shifts differing by
lattice  vectors have different spectra and are thus inequivalent.  Then
we review the concept of \textit{discrete torsion}, and clarify its relation to
models in which shifts differ by lattice vectors.

\subsection[Brother models]{Brother models}
\label{sec:brothermodel}

Let us start by clarifying under which conditions two models M and M$'$ are
equivalent. First, we restrict to the case without Wilson lines, where the
models M and M$'$ are described by the set of shifts $(V_1,V_2)$ and
$(V_1',V_2')$, respectively. Clearly, if the shifts are related by Weyl
reflections, i.e.\
\begin{equation}\label{eq:Weyl}
 (V_1',V_2')~=~(W\,V_1,W\,V_2)\;,
\end{equation}
where $W$ represents a series of Weyl reflections, one does obtain equivalent
models. Let us now turn to comparing the spectra of two models M and M$'$, where
\begin{equation}\label{eq:BrotherDef}
 (V_1',V_2')~=~(V_1+\Delta V_1,V_2+\Delta V_2)\;,
\end{equation}
with $\Delta V_1,\Delta V_2 \in \Lambda$. For future reference, we call models
related by equation~\eqref{eq:BrotherDef} `\textit{brother models}'. 

Brother models are also subject to modular invariance constraints. For the sake
of keeping the expressions simple, we restrict here to models fulfilling the
following (stronger) conditions:
\begin{subequations}\label{eq:strongmodularinv2}
\begin{eqnarray}
  V_{i}^{2} - v_{i}^{2} & = & 0 \mod 2  \quad (i = 1, 2)\;,\label{eq:strong1} \\
  V_{1} \cdot V_{2} - v_{1} \cdot v_{2} & = & 0 \mod 2\;.   \label{eq:strong2}
\end{eqnarray}
\end{subequations}
(In section~\ref{sec:GeneralizationDTAndBrothers} we will relax these
conditions.)
Equations \eqref{eq:strongmodularinv2} imply that $\Phi_\mathrm{vac}~=~1 $ in
the transformation phase \eqref{eq:transformationphase}. The condition that
$(V'_1,\,V'_2)$ fulfill  \eqref{eq:strongmodularinv2} leads to the following
constraints on $(\Delta V_1,\Delta V_2)$: 
\begin{subequations}\label{eq:modinvconditions}
\begin{eqnarray}
  V_{i} \cdot \Delta V_{i} & = & 0 \mod 1 \quad \quad i = 1, 2 \;,\\
  V_{1} \cdot \Delta V_{2} + \Delta V_{1} \cdot V_{2} + \Delta V_1\cdot\Delta V_2& = & 0 \mod 2\;.
\end{eqnarray}
\end{subequations}
Consider now the massless spectrum corresponding to the constructing
element
\begin{equation}
 g~=~(\theta^{k_1} \omega^{k_2}, n_\alpha e_\alpha)~\in~S
\end{equation}
of the models M and M$'$. For simplicity, we restrict our attention to
non-oscillator states. Physical states arise from tensoring together left- and
right-moving solutions of the masslessness condition
equation~\eqref{eq:MassEquation},
\begin{eqnarray}
|q + k_1 v_1 + k_2 v_2\rangle_\mathrm{R} \, \otimes\, |p + k_1 V_1 + k_2 V_2\rangle_\mathrm{L}  
&&\text{for M}\;,\\
|q + k_1 v_1 + k_2 v_2\rangle_\mathrm{R} \, \otimes\, |p' + k_1 V'_1 + k_2 V'_2\rangle_\mathrm{L} &&\text{for M}' \;,
\end{eqnarray}
where $p' = p - k_1\Delta V_1 - k_2 \Delta V_2$ and the shifted momenta of the
left-movers are identical for M and M$'$. According to equation~(\ref{eq:transformationphase}) with
$\Phi_\mathrm{vac}=1$, these massless states transform under the action of a commuting element
\begin{equation}
 h~=~(\theta^{t_1} \omega^{t_2}, m_\alpha e_\alpha) \in S \quad\text{with}\quad
 [h,g]~=~0
\end{equation}
with the phases
\begin{eqnarray}
 \Phi  & = & 
 e^{2\pi \I\, \left[(p + k_1 V_1 + k_2 V_2)\cdot (t_1 V_1 + t_2 V_2) - (q + k_1 v_1 + k_2 v_2)\cdot (t_1
     v_1 + t_2 v_2)\right]}
 \qquad\text{for M}\;,\nonumber\\
 \Phi' & = & e^{2\pi \I\, \left[(p' + k_1 V'_1 + k_2 V'_2)\cdot (t_1 V'_1 + t_2 V'_2) - (q + k_1 v_1 +
     k_2 v_2 )\cdot (t_1 v_1 + t_2 v_2)\right]}
 \ \quad\text{for M}'\;.\nonumber
\end{eqnarray}
By using the constraints~\eqref{eq:modinvconditions} and the properties of an integral lattice, 
$p\cdot \Delta V_i \in {\mathbb Z}$ for $p, \Delta V_i \in \Lambda$, the
mismatch between the phases can be simplified to
\begin{equation}
\Phi'~=~\Phi \, e^{-2\pi \I\, (k_1 t_2 - k_2 t_1)V_2\cdot\Delta V_1}\;.
\end{equation}
That is, the transformation phase of states in model M$'$ differs from the
transformation phase of states in model M by a relative phase
\begin{equation}\label{eq:brotherphase}
 \widetilde{\varepsilon}~=~e^{-2\pi \I (k_1 t_2 - k_2 t_1)V_2\cdot\Delta V_1}\;.
\end{equation}
According to the nomenclature `brother models', the relative phase
$\widetilde{\varepsilon}$ will be referred to as `brother phase'. It is
straightforward to see that the same relative phase occurs for oscillator
states, and the derivation can be repeated for shifts satisfying
\eqref{eq:newmodularinv} rather than \eqref{eq:strongmodularinv2}, yielding the
same qualitative result.

The (brother) phase $\widetilde{\varepsilon}$ has certain properties and the
fact that it can be non-trivial has important consequences. First of all,
$\widetilde{\varepsilon}$ depends on the definition of the model M$'$, i.e.\ on
the lattice vectors $(\Delta V_1,\Delta V_2)$. Furthermore, it clearly depends
on the constructing element $g$ and on the commuting element $h$, 
\begin{equation}
 \widetilde{\varepsilon}~=~ \widetilde{\varepsilon}(g,h)\;.
\end{equation}
It follows from the construction that the brother phase vanishes for $g =
(\mathbbm{1}, 0)$, i.e.\ for the untwisted sector. Thus the gauge group and the
untwisted sector coincide for brother models. On the other hand,
since the brother phase does not vanish in general, the brother models M and
M$'$ may have different twisted sectors, and therefore be inequivalent. This result extends also to the
case where we subject the shifts only to the weaker constraints \eqref{eq:newmodularinv}.

\subsubsection*{A $\boldsymbol{\mathbbm{Z}_3 \times \mathbbm{Z}_3}$ example}
\label{ex1}

Let us now study an example to illustrate the results obtained so far. Consider
a $\mathbbm{Z}_3 \times \mathbbm{Z}_3$ orbifold of $\text{E}_8 \times
\text{E}_8$ with standard embedding~\cite{Font:1989aj}, i.e.\ model M is defined by
\begin{equation}
 V_1~=~\frac{1}{3}\left(1,0,-1,0^{5} \right) \left( 0^{8} \right)
 \quad\text{and}\quad 
 V_2~=~\frac{1}{3}\left(0,1,-1,0^{5} \right) \left( 0^{8} \right)\;.
\end{equation}
The resulting model has an $\text{E}_6 \times \text{U}(1)^2 \times \text{E}_8$
gauge group, 84 $(\boldsymbol{\overline{27}},\boldsymbol{1})$ and 243 non-abelian
singlets with non-zero U(1) charges.\footnote{There are three additional singlets
$|q\rangle_\mathrm{R}\otimes\widetilde{\alpha}^i_{-1}|0\rangle_\mathrm{L}$ from the
10d SUGRA multiplet. All orbifold spectra are computed using~\cite{Orbifolder}.} Now define the brother model M$'$ by
\begin{equation}
 \Delta V_1~=~\left(0,-1,0,1,0^{4}\right) \left( 0^{8} \right)
 \quad\text{and}\quad 
 \Delta V_2~=~\left(1,0,0,0,1,0^{3}\right) \left( 0^{8} \right)\;,
\end{equation}
which fulfill the conditions~\eqref{eq:modinvconditions}.
From equation~\eqref{eq:brotherphase} we find the following non-trivial brother
phase
\begin{equation}\label{eq:brotherphasez3xz3}
 \widetilde{\varepsilon}(g,h)
 ~=~
 \widetilde{\varepsilon}(\theta^{k_1}\omega^{k_2},\theta^{t_1}\omega^{t_2})
 ~=~e^{\frac{2\pi\I}{3}\, (k_1\, t_2 - k_2\, t_1)}.
\end{equation}
As expected, the gauge group and the untwisted matter of model M$'$ remain the
same as in model M.
However, the twisted sectors get modified. The total number of generations is
reduced to 3 $(\boldsymbol{\overline{27}},\boldsymbol{1})$ and 27
$(\boldsymbol{27},\boldsymbol{1})$. The number of singlets remains the same as
before, but their localization properties change. 

Model M$'$ is not an unknown construction, but has been studied in the
literature in the context of $\mathbbm{Z}_3 \times \mathbbm{Z}_3$ orbifolds with
discrete torsion \cite{Font:1988mk}. As we shall see, the brother phase,
equation~\eqref{eq:brotherphasez3xz3}, is nothing but the discrete torsion phase (equation~(4) in
Ref.~\cite{Font:1988mk}). To make this statement more precise, we
briefly review discrete torsion in section \ref{sec:discretetorsionphase}, and analyze
its relation to the brother phase in section \ref{sec:comparison}.

\subsection[Discrete torsion phase for $\mathbbm{Z}_{N} \times \mathbbm{Z}_{M}$ orbifolds]{Discrete
  torsion phase for $\boldsymbol{\mathbbm{Z}_{N} \times \mathbbm{Z}_{M}}$ orbifolds}
\label{sec:discretetorsionphase}

Let us start with a brief review of discrete torsion in
orbifolds, following Vafa~\cite{Vafa:1986wx}.
The one-loop partition function $Z$ for a $\mathbbm{Z}_{N}\times \mathbbm{Z}_{M}$ 
orbifold has the overall structure
\begin{equation}
 Z~=~\sum_{\substack{g,h \\ [g,h]=0}} \varepsilon (g,h)\, Z(g,h)\;,
\end{equation}
where the sum runs over pairs of commuting space group elements $g,h\in
\mbox{S}$ and the $\varepsilon (g,h)$ are relative phases between the different
terms in the partition function and thus between the different sectors.
Different assignments of phases lead, in general, to different orbifold models.

Modular invariance strongly constrains the torsion phases
\cite{Vafa:1986wx}: 
\begin{subequations}\label{eq:phaseconstraints}
\begin{eqnarray}
 \varepsilon(g_{1} g_{2},g_{3}) &=&
 \varepsilon(g_{1},g_{3})\, \varepsilon(g_{2},g_{3})\;,
 \label{eq:phaseconstraints1} \\
 \varepsilon(g_{1},g_{2}) &=&
 \varepsilon(g_{2},g_{1})^{-1}\;. \label{eq:phaseconstraints2} 
\end{eqnarray} 
Further, we use the convention
\begin{equation}
 \label{eq:phaseconstraints3} \varepsilon(g,g)
 ~=~ 1 \; . 
\end{equation} 

At two--loop, the partition function allows to switch on analogous phases,
$\varepsilon(g_{1},h_{1};g_{2},h_{2})$. From the requirement of factorizability
of the two--loop partition function one infers \cite{Vafa:1986wx} 
\begin{equation}
 \varepsilon(g_{1},h_{1};g_{2},h_{2})
 ~=~\varepsilon(g_{1},h_{1})\,\varepsilon(g_{2},h_{2})\;.
\end{equation}
\end{subequations} 

Following the discussion of Ref.~\cite{Font:1988mk}, in orbifolds without Wilson lines $g,h$ are chosen to be
elements of the point group $P$. In $\mathbbm{Z}_N$ orbifolds, due to this choice and 
equations~\eqref{eq:phaseconstraints} the phases have to be trivial,
\begin{equation}
\varepsilon(g,h)~=~1 \qquad \forall g,h \in P\;.
\end{equation}
Therefore, in the case of $\mathbbm{Z}_{N}$ orbifolds without Wilson lines,
non-trivial discrete torsion cannot be introduced.

In $\mathbbm{Z}_{N}\times \mathbbm{Z}_{M}$ orbifolds, still without Wilson
lines, the situation is different because there are independent pairs of
elements (such that the first element is not a power of the second) which
commute with each other.  If we take two point group elements
$g=\theta^{k_1}\omega^{k_2}$ and
$h=\theta^{t_1}\omega^{t_2}$, the equations~\eqref{eq:phaseconstraints}
determine the shape of the corresponding phase, 
\begin{equation}\label{eq:torsionphase2}
 \varepsilon(g,h)
 ~=~
 \varepsilon(\theta^{k_1}\omega^{k_2},\theta^{t_1}\omega^{t_2})
 ~=~
 e^{\frac{2\pi\I\, m}{M}(k_1 t_2-k_2 t_1)}\,\,,
\end{equation}
where $m\in \mathbbm{Z}$ \cite{Font:1988mk}. In particular, there are only $M$
inequivalent assignments of $\varepsilon$.

The most important consequence of non-trivial $\varepsilon$-phases for our
discussion is that they modify the boundary conditions for twisted states and
thus change the twisted spectrum. This can be seen from the transformation phase
of equation~(\ref{eq:transformationphase}), which is modified in the presence of
discrete torsion according to 
\begin{equation}\label{eq:modifiedtransformationphase}
 \Phi \longmapsto \varepsilon(g,h) \,\Phi\;.
\end{equation}

\subsection{Brother models versus discrete torsion}
\label{sec:comparison}

\begin{figure}[!t]
\centerline{\includegraphics{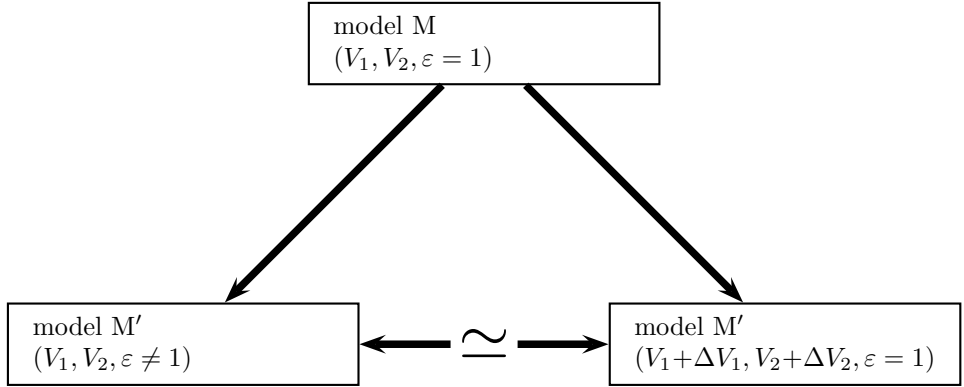}}
\caption{Models with non-trivial discrete torsion have an equivalent
description as models with trivial discrete torsion but a different gauge
embedding.}
\label{fig:MirageTorsion}
\end{figure}

Let us now come back to the task of establishing the relation between the
discrete torsion phase and the brother phase as introduced in section
\ref{sec:brothermodel}. From equations~\eqref{eq:brotherphase} and~\eqref{eq:torsionphase2} it is clear
that both phases can be made to coincide.
More precisely, since $V_2$ can be written as $V_{2}=\frac{\lambda_{2}}{M}$ with 
$\lambda_{2} \in \Lambda$ (cf.\ equation~\eqref{eq:latticeconditions}), one can
achieve
\begin{equation}\label{eq:mdef}
 -V_{2} \cdot \Delta V_{1}
 ~=~
 \frac{m}{M} 
\end{equation}
for an appropriate choice of $\Delta V_1\in\Lambda$. Since the solutions to the
mass equations and the projection conditions are the same in a model with
discrete torsion and a brother model, whose associated phases fulfill equation~\eqref{eq:mdef}, the spectra of
both models coincide. We will therefore regard both models as \emph{equivalent}.
This means that introducing a discrete torsion phase, equation~\eqref{eq:torsionphase2}, is equivalent to
changing the gauge embedding according to 
\begin{equation}
 (V_1,V_2) ~ \to ~ (V_1+\Delta V_1,V_2+\Delta V_2)
\end{equation}
with $\Delta V_i\in\Lambda$ and $-V_{2} \cdot \Delta V_{1}=m/M$. In particular,
the assignment of discrete torsion to a given $\mathbbm{Z}_N\times\mathbbm{Z}_M$
model is a `gauge-dependent' statement in the sense that torsion can be traded
for changing the gauge embedding (cf.\ Fig.~\ref{fig:MirageTorsion}).

To illustrate our result, we construct the standard embedding models for
$\mathbbm{Z}_N \times \mathbbm{Z}_M$ orbifolds with an $\mathrm{E}_8 \times
\mathrm{E}_8$ lattice of Ref.~\cite{Font:1988mk} with discrete torsion in terms of
non-standard embedding shifts without discrete torsion (brother models). We use
the following recipe to construct brother models, i.e.\ mimic models with
discrete torsion:\\ 
For a given set of shifts $V_1$ and $V_2$ fulfilling the modular invariance conditions,
find a new set of shifts $V_1'=V_1+\Delta V_1$ and $V_2'=V_2+\Delta V_2$
with the following properties:
\renewcommand{\labelenumi}{(\roman{enumi})}
\begin{enumerate}
 \item the new shifts differ from the original set only by lattice vectors,
  i.e.\ $\Delta V_1,\Delta V_2\in\Lambda$
 \item the new shifts also fulfill the modular invariance conditions,
  and
 \item the `interference term' $V_2\cdot \Delta V_1$ is not an integer.
\end{enumerate}
In practice (and for any $N,M$), the above properties can be expressed in
terms of linear Diophantine equations for which we always find solutions.

Possible choices for the shifts $(V_1+\Delta V_1,V_2+\Delta V_2)$ are shown in
Tab.~\ref{tab:DiscreteTorsionBrothers}, where we list the shifts of torsionless
models equivalent to the discrete torsion model of Ref.~\cite{Font:1988mk}.

\begin{table}[!h]
\begin{center}
\begin{tabular}{|l|r|l|l|}
\hline
orbifold & torsion $\varepsilon$ & shift $V_1$ & shift $V_2$\\
\hline\hline
$\Z2\times\Z2$ &
$1$ &
$\displaystyle \left(\tfrac{1}{2},0,-\tfrac{1}{2},0,0,0,0,0\right)$ & 
$\displaystyle \left(0,\tfrac{1}{2},-\tfrac{1}{2},0 ,0, 0, 0, 0\right)$\\
& $-1$ &
$\displaystyle \left(\tfrac{1}{2}, -1,-\tfrac{1}{2},1,0,0,0,0\right)$ & 
$\displaystyle \left(1,\tfrac{1}{2},-\tfrac{1}{2},0,1,0,0,0\right)$\\
\hline
$\Z4\times\Z2$ &
$1$ &
$\displaystyle \left(\tfrac{1}{4},0,-\tfrac{1}{4},0,0,0,0,0\right)$ & 
$\displaystyle \left(0,\tfrac{1}{2},-\tfrac{1}{2},0,0,0,0,0\right)$ \\
& $-1$ &
$\displaystyle \left(\tfrac{1}{4},-1,-\tfrac{1}{4},1,0,0,0,0\right)$ & 
$\displaystyle \left(2,\tfrac{1}{2},-\tfrac{1}{2},0,0,0,0,0\right)$ \\
\hline
$\Z6\times\Z2$ &
$1$ &
$\displaystyle \left(\tfrac{1}{6},0,-\tfrac{1}{6},0,0,0,0,0\right)$ & 
$\displaystyle \left(0,\tfrac{1}{2},-\tfrac{1}{2},0 ,0, 0, 0, 0\right)$\\
& $-1$ &
$\displaystyle \left(\tfrac{1}{6},-1,-\tfrac{1}{6},1,0,0,0,0\right)$ & 
$\displaystyle \left(3,\tfrac{1}{2},-\tfrac{1}{2},0,1,0,0,0\right)$ \\
\hline
$\Z6'\times\Z2$ &
$1$ &
$\displaystyle \left(\tfrac{1}{6},\tfrac{1}{6},-\tfrac{1}{3},0,0,0,0,0\right)$ & 
$\displaystyle \left(\tfrac{1}{2},0,-\tfrac{1}{2},0 ,0, 0, 0, 0\right)$\\
& $-1$ &
$\displaystyle \left(-\tfrac{5}{6},\tfrac{7}{6},-\tfrac{1}{3},1,1,0,0,0\right)$ & 
$\displaystyle \left(\tfrac{1}{2},3,-\tfrac{1}{2},1,0,0,0,0\right)$\\
\hline
$\Z3\times\Z3$ &
$1$ &
$\displaystyle \left(\tfrac{1}{3},0,-\tfrac{1}{3},0,0,0,0,0\right)$ & 
$\displaystyle \left(0,\tfrac{1}{3},-\tfrac{1}{3},0 ,0, 0, 0, 0\right)$\\
& $\mathrm{e}^{2\pi\I\tfrac{1}{3}}$ &
$\displaystyle \left(\tfrac{1}{3},-1,-\tfrac{1}{3},1,0,0,0,0\right)$ & 
$\displaystyle \left(1,\tfrac{1}{3},-\tfrac{1}{3},0,1,0,0,0\right)$ \\
& $\mathrm{e}^{2\pi\I\tfrac{2}{3}}$ &
$\displaystyle \left(\tfrac{1}{3},-2,-\tfrac{1}{3},0,0,0,0,0\right)$ & 
$\displaystyle \left(2,\tfrac{1}{3},-\tfrac{1}{3},0,0,0,0,0\right)$ \\
\hline
$\Z6\times\Z3$ &
$1$ &
$\displaystyle \left(\tfrac{1}{6},0,-\tfrac{1}{6},0,0,0,0,0\right)$ & 
$\displaystyle \left(0,\tfrac{1}{3},-\tfrac{1}{3},0 ,0, 0, 0, 0\right)$\\
& $\mathrm{e}^{2\pi\I\tfrac{1}{3}}$ &
$\displaystyle \left(\tfrac{1}{6},-1,-\tfrac{1}{6},1,0,0,0,0\right)$ & 
$\displaystyle \left(2,\tfrac{1}{3},-\tfrac{1}{3},0,0, 0, 0, 0\right)$\\
& $\mathrm{e}^{2\pi\I\tfrac{2}{3}}$ &
$\displaystyle \left(\tfrac{1}{6},-2,-\tfrac{1}{6},0,0,0,0,0\right)$ & 
$\displaystyle \left(4,\tfrac{1}{3},-\tfrac{1}{3},0,0, 0, 0, 0\right)$\\
\hline
$\Z4\times\Z4$ &
$1$ &
$\displaystyle \left(\tfrac{1}{4},0,-\tfrac{1}{4},0,0,0,0,0\right)$ & 
$\displaystyle \left(0,\tfrac{1}{4},-\tfrac{1}{4},0,0,0,0,0\right)$\\
& $\I$ &
$\displaystyle \left(\tfrac{1}{4},-1,-\tfrac{1}{4},1,0,0,0,0\right)$ &
$\displaystyle \left(1,\tfrac{1}{4},-\tfrac{1}{4},0,1,0,0,0\right)$\\ 
& $-1$ &
$\displaystyle \left(\tfrac{1}{4},-2,-\tfrac{1}{4},0,0,0,0,0\right)$ &
$\displaystyle \left(2,\tfrac{1}{4},-\tfrac{1}{4},0,0,0,0,0\right)$\\ 
& $-\I$ &
$\displaystyle \left(\tfrac{1}{4},-3,-\tfrac{1}{4},1,0,0,0,0\right)$ &
$\displaystyle \left(3,\tfrac{1}{4},-\tfrac{1}{4},0,1,0,0,0\right)$\\ 
\hline
$\Z6\times\Z6$ &
$1$ &
$\displaystyle \left(\tfrac{1}{6},0,-\tfrac{1}{6},0,0,0,0,0\right)$ & 
$\displaystyle \left(0,\tfrac{1}{6},-\tfrac{1}{6},0,0,0,0,0\right)$\\
& $\mathrm{e}^{2\pi\I\tfrac{1}{6}}$ &
$\displaystyle \left(\tfrac{1}{6},-1,-\tfrac{1}{6},1,0,0,0,0\right)$ & 
$\displaystyle \left(1,\tfrac{1}{6},-\tfrac{1}{6},0,1,0,0,0\right)$\\
& $\mathrm{e}^{2\pi\I\tfrac{1}{3}}$ &
$\displaystyle \left(\tfrac{1}{6},-2,-\tfrac{1}{6},0,0,0,0,0\right)$ & 
$\displaystyle \left(2,\tfrac{1}{6},-\tfrac{1}{6},0,0,0,0,0\right)$\\
& $-1$ &
$\displaystyle \left(\tfrac{1}{6},-3,-\tfrac{1}{6},1,0,0,0,0\right)$ & 
$\displaystyle \left(3,\tfrac{1}{6},-\tfrac{1}{6},0,1,0,0,0\right)$\\
& $\mathrm{e}^{2\pi\I\tfrac{2}{3}}$ &
$\displaystyle \left(\tfrac{1}{6},-4,-\tfrac{1}{6},0,0,0,0,0\right)$ & 
$\displaystyle \left(4,\tfrac{1}{6},-\tfrac{1}{6},0,0,0,0,0\right)$\\
& $\mathrm{e}^{2\pi\I\tfrac{5}{6}}$ &
$\displaystyle \left(\tfrac{1}{6},-5,-\tfrac{1}{6},1,0,0,0,0\right)$ & 
$\displaystyle \left(5,\tfrac{1}{6},-\tfrac{1}{6},0,1,0,0,0\right)$\\
\hline
\end{tabular}
\caption{$\Z{N}\times\Z{M}$ models with discrete torsion and standard embedding are equivalent to
  models without discrete torsion and non-standard embedding. We write the torsion phase factor as
  $\varepsilon=\mathrm{e}^{-2\pi\I\, V_2\cdot\Delta V_1}$. The components of the shifts within the second
  $\E8$ all vanish. This result also applies to orbifold models in SO(32).}
\label{tab:DiscreteTorsionBrothers}
\end{center}
\end{table}

Our result has important consequences for the classification of
$\mathbbm{Z}_N\times\mathbbm{Z}_M$ orbifolds. Introducing a discrete torsion
phase in the sense of Ref.~\cite{Font:1988mk} does not lead to new models. That
is, all models with this discrete torsion can be equivalently obtained by
scanning over torsionless models only. This will be important for our
classification in section~\ref{sec:ClassifZ3xZ3}.

It is also instructive to interpret the equivalence between discrete torsion and
changing the gauge embedding in terms of geometry. Discrete torsion can be
regarded as a property of the 6D compact space while changing the gauge
embedding affects the (left-moving) coordinates of the gauge lattice only. Hence
one might argue that discrete torsion and choosing a different gauge embedding
are two different features of orthogonal dimensions. However, by embedding the
`spatial' twist in the gauge degrees of freedom, these features get combined in
such a way that it is no longer possible to make a clear separation. Using a
more technical language one might rephrase this statement by saying that, since
physical states arise from tensoring left- and right-movers together, the phases
$\varepsilon$ and $\widetilde{\varepsilon}$ cannot be distinguished.
Consequently, properties of the zero-modes cannot be ascribed neither to the gauge
embedding alone nor to the presence of discrete torsion, but only to both.

\section{Generalized discrete torsion}
\label{sec:GeneralizationDTAndBrothers}

The results of the previous section can be generalized. To see this, we first 
generalize the brother phase of section~\ref{sec:brothermodel} for orbifolds
with Wilson lines. In a second step, we compare the emerging phases to what is
known as generalized discrete torsion \cite{Gaberdiel:2004vx}. As before, we can
relate both phases.

\subsection{Generalized brother models}
\label{sec:GeneralizedBrothers}

Let us turn to the discussion of orbifolds with Wilson lines~\cite{Ibanez:1986tp}. A (torsionless) model
M is defined by $(V_1,V_2,A_\alpha)$. A brother model M$'$ appears by adding lattice vectors to
the shifts and Wilson lines, i.e.\ M$'$ is defined by 
\begin{equation}
 (V'_1, V'_2, A'_\alpha)~=~(V_1+\Delta V_1, V_2+\Delta V_2, A_\alpha+\Delta A_\alpha)\;,
\end{equation}
with $\Delta V_i,\Delta A_\alpha \in \Lambda$. From the
conditions~\eqref{eq:newmodularinv}, the choice of lattice vectors $(\Delta
V_i,\Delta A_\alpha)$ is constrained by
\begin{subequations}\label{eq:newModularInvforDelta}
\begin{eqnarray}
M \left(V_1\cdot\Delta V_2 + V_2\cdot\Delta V_1 + \Delta V_1 \cdot\Delta V_2 \right) & = & 0\text{ mod }
2~\equiv~ 2\,x \;,\label{eq:newModularInvforDeltaVV}\\
N_\alpha \left(V_i\cdot\Delta A_\alpha + A_\alpha\cdot\Delta V_i + \Delta V_i \cdot\Delta A_\alpha
\right) & = &  0\text{ mod }
2~\equiv~ 2\,y_{i\alpha} \;,\label{eq:newModularInvforDeltaVA}\\
Q_{\alpha\beta} \left(A_\alpha\cdot\Delta A_\beta + A_\beta\cdot\Delta A_\alpha + \Delta A_\alpha\cdot\Delta A_\beta
\right) & = &  0\text{ mod }
2~\equiv~ 2\,z_{\alpha\beta}\;,\label{eq:newModularInvforDeltaAA}
\end{eqnarray}
\end{subequations}
where $x,\,y_{i\alpha},\,z_{\alpha\beta}\,\in{\mathbb Z}$.

Repeating the steps of section~\ref{sec:brothermodel} one arrives at a
`generalized brother phase'
\begin{eqnarray}
 \widetilde{\varepsilon}&=&\exp\left\{-2\pi \I\,\left[ 
   (k_1\, t_2-k_2\, t_1)\left(V_{2} \cdot \Delta V_{1} - \frac{x}{M}\right) \phantom{\frac{z_{\alpha\beta}}{Q_{\alpha\beta}}}\right.\right.\nonumber\\*
 & & \hphantom{\exp\left\{\right.}{}\left.\left.
 + (k_1\, m_{\alpha} - t_1\, n_{\alpha})\left(A_{\alpha} \cdot \Delta V_{1}-\frac{y_{1\alpha}}{N_\alpha}\right)\right.\right.
 \nonumber\\*
 & &  \hphantom{\exp\left\{\right.}{}\left.\left.
  + (k_2\, m_{\alpha}-t_2\, n_{\alpha})\left(A_{\alpha} \cdot \Delta
    V_{2}-\frac{y_{2\alpha}}{N_\alpha}\right) \right.\right.
 \nonumber\\*
 & &  \hphantom{\exp\left\{\right.}{}\left.\left.
  + n_{\alpha}\, m_{\beta}\left( A_{\beta} \cdot \Delta
    A_{\alpha}-\frac{z_{\alpha\beta}}{Q_{\alpha\beta}}\right)\right]
 \right\}\;,
 \label{eq:generalbrotherphase} 
\end{eqnarray}
corresponding to the constructing element
$g=(\theta^{k_1}\omega^{k_2},n_{\alpha}e_{\alpha})$ and the commuting element
$h=(\theta^{t_1}\omega^{t_2},m_{\alpha}e_{\alpha})$. One can see that $D_{\alpha\beta}\equiv A_\beta\cdot\Delta
A_\alpha - z_{\alpha\beta}/Q_{\alpha\beta}$ is (almost) antisymmetric in $\alpha,\,\beta$,
\begin{equation}
D_{\alpha\beta} = -D_{\beta\alpha}\text{ mod }1\,.
\end{equation}

Notice that also in the case of orbifolds with lattice-valued Wilson lines,
$A_\alpha\in\Lambda$, the last three terms of
equation~\eqref{eq:generalbrotherphase} can be non-trivial, giving rise to new
brother models.

\subsubsection*{Brother models in $\boldsymbol{\mathbbm{Z}_N}$ orbifolds}

From equation~\eqref{eq:generalbrotherphase}, it is clear that the generalized brother phase is also
important for $\mathbbm{Z}_N$ orbifolds. More precisely, in $\mathbbm{Z}_N$ orbifolds with Wilson lines,
the second and fourth lines of equation~\eqref{eq:generalbrotherphase} are not always trivial and thus
also lead to brother models. 

Let us illustrate this with an example in $\mathbbm{Z}_4$ with twist
$v=\frac{1}{4}(-2,\,1,\,1;\,0)$ acting on the compactification lattice $\Gamma=$
SO(4)$^3$, and standard embedding~\cite{Katsuki:1990bf}. 
\label{ex2}
The gauge group is \E{6}$\times$SU(2)$\times$\E{8}. By turning on the
lattice-valued Wilson lines 
\begin{equation}
 A_1=\left(0^8 \right)\left(1^2,0^6\right), \quad\quad\quad A_5=A_6=\left(0^8 \right)\left(0,1^2,0^5\right),
\end{equation}
a non-trivial generalized brother phase with $D_{15}=D_{16}=-\frac{1}{2}$ is introduced. The untwisted
and first twisted sectors remain unchanged, but the number of (anti-) families in
the second twisted sector is reduced from 10 $(\boldsymbol{\overline{27}},\boldsymbol{1},\boldsymbol{1})$
+ 6 $(\boldsymbol{27},\boldsymbol{1},\boldsymbol{1})$ to 6 $(\boldsymbol{\overline{27}},\boldsymbol{1},\boldsymbol{1})$
+ 2 $(\boldsymbol{27},\boldsymbol{1},\boldsymbol{1})$.

\subsection{Generalized discrete torsion}
\label{sec:GeneralizedDiscreteTorsion}

In section~\ref{sec:discretetorsionphase} we have discussed the discrete torsion
phase as introduced in Ref.~\cite{Font:1988mk}. More recently, this concept has
been extended by introducing a generalized discrete torsion phase in the context
of type IIA/B string theory~\cite{Gaberdiel:2004vx}. This generalized torsion
phase depends on the fixed points of the orbifold. It weights differently terms
in the partition function corresponding to the same twisted sector but different
fixed points, and is constrained by modular invariance. 

Following the steps of section~\ref{sec:discretetorsionphase} and considering
$g,h\in S$, we write down the general solution of
equations~\eqref{eq:phaseconstraints} for the discrete torsion phase
as\footnote{Note that we employ the stronger
constraints~\eqref{eq:phaseconstraints} rather than the conditions presented
in~\cite{Gaberdiel:2004vx}. It might be possible to relax
condition~\eqref{eq:phaseconstraints2}, in which case additional possibilities
could arise. We ignore this possibility in the present study.}
\begin{equation}\label{eq:generalizedtorsionphase}
\varepsilon(g,h)~=~e^{2 \pi \I\,[a\, (k_1\, t_2 - k_2\, t_1) 
+ b_{\alpha}\, (k_1\, m_{\alpha} - t_1\, n_{\alpha}) 
+ c_{\alpha}\, (k_2\, m_{\alpha} - t_2\, n_{\alpha}) 
+ d_{\alpha \beta}\, n_{\alpha}\, m_{\beta}]}\;.
\end{equation}
Modular invariance constrains the values of $a,b_{\alpha},c_{\alpha},d_{\alpha
\beta}$. Therefore $a=\widetilde{a}/M,\,b_\alpha =
\widetilde{b}_\alpha/N_\alpha$, $c_\alpha = \widetilde{c}_\alpha/N_\alpha$,
$d_{\alpha\beta}=\widetilde{d}_{\alpha\beta}/N_{\alpha\beta}$ with
$\widetilde{a},\,\widetilde{b}_\alpha$, $\widetilde{c}_\alpha$,
$\widetilde{d}_{\alpha\beta} \in \mathbbm{Z}$, $N_{\alpha\beta}$ being the
greatest common divisor of $N_\alpha$ and $N_\beta$. In addition, $d_{\alpha
\beta}$ must be antisymmetric in $\alpha,\,\beta$.

The parameters $b_\alpha$, $c_\alpha$, $d_{\alpha\beta}$ are additionally
constrained by the geometry of the orbifold. It is not hard to see that if
$e_\alpha \simeq e_\beta$ on the orbifold, then $b_\alpha=b_\beta,\,
c_\alpha=c_\beta$ and $d_{\alpha\beta}=0$ must hold (cf.\ the examples below).

The generalized discrete torsion is not restricted only to $\mathbbm{Z}_N\times\mathbbm{Z}_M$
orbifolds, as the usual discrete torsion was, but will likewise appear in the $\mathbbm{Z}_N$
case. Clearly, since in $\mathbbm{Z}_N$ orbifolds there is only one shift, the parameters $a$ and
$c_\alpha$ vanish.

\subsubsection*{Examples}
\label{ex3}

Let us consider the $\mathbbm{Z}_3\times \mathbbm{Z}_3$ orbifold compactified on
an $\SU3^3$ lattice. In this case we have $e_1 \simeq e_2,\, e_3 \simeq e_4$
and $e_5 \simeq e_6$ on the orbifold. This implies that there are only three
independent $b_\alpha$, namely $b_1,\,b_3,\,b_5$, while $b_2 = b_1,\,b_4 =
b_3,\,b_6 = b_5$. Analogously, only $c_1,\,c_3,\,c_5$ are independent. Further,
the antisymmetric matrix $d_{\alpha\beta}$ takes the form
\begin{equation}
 d_{\alpha\beta}~=~ \left(
 \begin{array}{cccccc}
 0&0&d_1&d_1&d_2&d_2\\
 0&0&d_1&d_1&d_2&d_2\\
 -d_1&-d_1&0&0&d_3&d_3\\
 -d_1&-d_1&0&0&d_3&d_3\\
 -d_2&-d_2&-d_3&-d_3&0&0\\
 -d_2&-d_2&-d_3&-d_3&0&0
 \end{array}
 \right)\;.
 \label{eq:dmatrixZ3xZ3}
\end{equation}
Including the parameter $a$, there are 10 independent discrete torsion
parameters, which can take values $0$, $\tfrac{1}{3}$ or $\tfrac{2}{3}$.

For the $\mathbbm{Z}_2 \times \mathbbm{Z}_2$ orbifold on an $\SU2^6$ lattice an
analogous consideration shows that there are no restrictions for the discrete
torsion parameters. Therefore, there are $1+6+6+15=28$ independent parameters
$a,\,b_\alpha,\,c_\alpha,\,d_{\alpha\beta}$, with values either 0 or
$\tfrac{1}{2}$. However, since the coefficients $n_\alpha m_\beta$ of
$d_{\alpha\beta}$ for $(\alpha,\,\beta)\in\{(1,2),\,(3,4),\,(5,6)\}$ vanish, the
corresponding $d_{\alpha\beta}$ are not physical, leading to 25 effective
parameters.
 
\subsubsection*{Generalized discrete torsion and local spectra}

In order to understand the action of the generalized discrete torsion, let us
consider the following example. 
\label{ex4}
We start with the $\mathbbm{Z}_3\times\mathbbm{Z}_3$ standard embedding without
Wilson lines, $A_\alpha=0$, and switch on the discrete torsion phase,
equation~\eqref{eq:generalizedtorsionphase}, with $b_3=b_4=\frac{1}{3}$. The
total number of families is reduced from 84
$(\boldsymbol{\overline{27}},\boldsymbol{1})$ to 24
$(\boldsymbol{\overline{27}},\boldsymbol{1})$ and 12
$(\boldsymbol{27},\boldsymbol{1})$. 

Due to its form, the discrete torsion phase
$\varepsilon=e^{2\pi\I\,b_{\alpha}\,(k_1\, m_{\alpha}-t_1\,n_\alpha)}$
distinguishes between different fixed points of a particular twisted sector.
That is, generalized discrete torsion can be thought of as a local feature.  In
general, the additional phase at a given fixed point coincides with a brother
phase of the torsionless model (cf.\ first term of
equation~\eqref{eq:generalbrotherphase}), i.e.\ locally one can find $\Delta
V_i$ such that
\begin{equation}\label{eq:epsilonloc}
 \varepsilon~=~e^{2\pi\I\,b_{\alpha}\,(k_1\, m_{\alpha}-t_1\,n_\alpha)}
 ~=~e^{-2\pi\I\,(k_1\, t_2-k_2\, t_1) 
               \left(V_2\cdot\Delta V_1 - \frac{x}{3}\right)}
\end{equation}
with appropriate $x$.
Then, each local spectrum coincides with the local spectrum of some brother
model. The interpretation of generalized discrete torsion in terms of
`localized discrete torsion' parallels the concept of local shifts (cf.\
\cite{Buchmuller:2004hv,Buchmuller:2006ik}) in orbifolds with Wilson lines.

Note that $\Delta V_i$ as in \eqref{eq:epsilonloc} cannot be found for twisted
sectors where $b_\alpha$ corresponds to a direction $e_\alpha$ of a fixed torus,
where $b_\alpha$ projects out all states of the sector.

\begin{figure}[!t!]
\centerline{\subfigure[{}]{\includegraphics{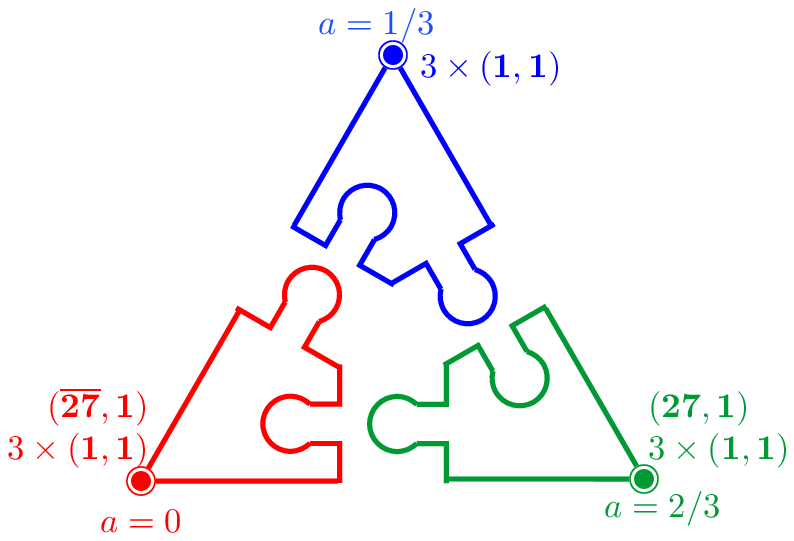}}
\qquad
\subfigure[{}]{\includegraphics{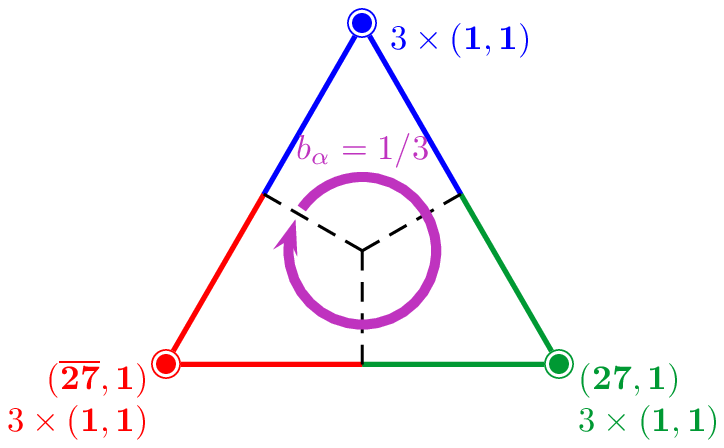}}}
\caption{Sketch of a (2D) \SU3 plane of a $\mathbbm{Z}_3\times\mathbbm{Z}_3$
orbifold (the second plane in the example). Parts (`corners') from different
brother models (a) can be `sewed together' to a model in which the torsion phase
differs for different fixed points. This is equivalent to switching on the
generalized discrete torsion phase $b_\alpha$ (b).}
\label{fig:puzzle}
\end{figure}

For concreteness, we first focus on the three fixed points in the second torus of the $T_{(0,1)}$
twisted sector. As depicted in Fig.~\ref{fig:puzzle}, the local spectra of the three brother
models, $a\equiv -\left(V_2\cdot\Delta V_1 - \frac{x}{3}\right)=0,\,\frac{1}{3},\,\frac{2}{3}$, can be combined consistently into one model with
$b_3=b_4=\frac{1}{3}$. On the other hand, in the $T_{(1,0)}$
twisted sector there is a fixed torus in the directions $e_3,\,e_4$; thus the sector is empty.

This procedure can also be applied to the terms $c_\alpha$ and $d_{\alpha\beta}$ of the generalized
discrete torsion phase, equation~\eqref{eq:generalizedtorsionphase}.

\subsubsection*{Generalized brother models versus generalized discrete torsion}

As in our previous discussion in chapter 3, also the generalized versions of the
discrete torsion phase and the brother phase have a very similar form. Indeed,
whenever there are non-trivial solutions to
equations~\eqref{eq:newModularInvforDelta}, one can equivalently describe models
with generalized discrete torsion phase in terms of generalized brother models.
This is the generic case.

However, there are exceptions. Namely, as we will explain below, models with
$d_{\alpha\beta}\neq 0$ in $\mathbbm{Z}_3\times\mathbbm{Z}_3$ orbifolds without Wilson lines cannot be
interpreted in terms of brother models.

Consider the fourth part of the generalized discrete torsion phase of
equation~\eqref{eq:generalizedtorsionphase},
\begin{equation}\label{eq:dtermdiscretetorsion}
\varepsilon~=~e^{2\pi\I\,d_{\alpha\beta}\,n_{\alpha} m_{\beta}}\;,
\end{equation}
with $d_{\alpha\beta}\in \left\lbrace 0,\tfrac{1}{3},\tfrac{2}{3}\right\rbrace$. An analogous
term appears in the generalized brother phase as
\begin{equation}\label{eq:dtermbrotherphase}
\widetilde{\varepsilon}~=~\exp\left[-2 \pi \I\, n_{\alpha}m_{\beta}\left(A_\beta \cdot \Delta
    A_{\alpha}-\frac{z_{\alpha\beta}}{Q_{\alpha\beta}} \right)\right]\;,
\end{equation}
where $Q_{\alpha\beta}~=~3$, since the Wilson lines have order 3. In general, both phases can be made
coincide by choosing $\Delta A_{\alpha}\in \Lambda$ such that 
\begin{equation}
-\left(A_\beta \cdot \Delta A_{\alpha}-\frac{z_{\alpha\beta}}{3} \right)=d_{\alpha\beta}\;.
\end{equation}
On the other hand, in the case when $A_\alpha=0$ and $\Delta A_\alpha\neq 0$,
equation~\eqref{eq:dtermbrotherphase} simplifies to 
\begin{equation}\label{eq:dtermbrotherphaseZ3xZ3}
 \widetilde{\varepsilon}
 ~=~e^{2\pi \I\, n_{\alpha}m_{\beta}\left(\frac{z_{\alpha\beta}}{3}\right)}
 ~=~e^{2\pi \I\, n_{\alpha}m_{\beta}\left(\frac{\Delta A_\alpha\cdot\Delta A_\beta}{2} \right)}\;,
\end{equation}
where the second equality follows from the definition of $z_{\alpha\beta}$,
equation~\eqref{eq:newModularInvforDeltaAA}. As $\Delta A_\alpha$
are lattice vectors, this equality can only hold if $z_{\alpha\beta}=0\text{ mod }3$, which implies
that the brother phase equation~\eqref{eq:dtermbrotherphaseZ3xZ3} is trivial. Thus, in this case, the
generalized discrete torsion phase leads to models which cannot arise by adding lattice vectors to shifts
and Wilson lines.

In summary, the generalized discrete torsion phases admit more possible
assignments than the generalized brother phases. Nevertheless, a large class of
the models with generalized discrete torsion has an equivalent description in
terms of models with a modified gauge embedding.

Our results have important implications. By introducing generalized discrete
torsion, or lattice-valued Wilson lines, one can control the local spectra. We
therefore expect that introducing generalized discrete torsion, or alternatively
shifting the Wilson lines by lattice vectors, will gain a similar importance as
discrete Wilson lines~\cite{Ibanez:1986tp} for orbifold model building.

As stated above, switching on generalized discrete torsion can lead to the
disappearance of complete local spectra. This raises the question of how to
interpret this fact in terms of geometry. Some of the localized zero-modes can
be viewed as blow-up modes which allow to resolve the orbifold singularity
associated to a given fixed point
\cite{Walton:1988bu,Aspinwall:1994ev,Vafa:1994rv} (see
\cite{Lust:2006zh,Honecker:2006qz,Nibbelink:2007rd} for recent developments). If
at a given fixed point there are no zero modes, one might argue that, therefore,
the associated singularity cannot be `blown up'. In what follows, we shall
advertise an alternative interpretation.

\subsection{Connection to non-factorizable orbifolds}
\label{sec:Z2xZ2}

We find that in many cases orbifold models M with certain geometry, i.e.\
compactification lattice $\Gamma$, and generalized discrete torsion switched on
are equivalent to torsionless models M$'$ based on a different lattice
$\Gamma'$. Model M$'$ has less fixed points than M, and the mismatch turns out
to constitute precisely the `empty' fixed points of model M.

The simplest examples are based on $\mathbbm{Z}_2\times\mathbbm{Z}_2$ orbifolds
with standard embedding and without Wilson lines.\label{ex5}
As compactification lattice $\Gamma$, we choose an $\SU2^6$
lattice~\cite{Forste:2004ie}. As we have seen in
section~\ref{sec:GeneralizedDiscreteTorsion}, in this case there are 25 physical
parameters for generalized discrete torsion, with values either 0 or
$\frac{1}{2}$. For concreteness, we restrict to the 12 $d_{\alpha\beta}$
parameters and scan over all $2^{12}$ models.

\begin{table}[b!]
\def\oldwidth{\captionwidth}
\setlength{\captionwidth}{0.85\textwidth} 
{\small{
\begin{center}
\begin{tabular}{|c||c|c|c|c|r||c||l|}
\hline
& $T_{(1,0)}$  & $T_{(0,1)}$  & $T_{(1,1)}$  & total            & $\#_S$ & $d_{\alpha\beta} = \tfrac{1}{2}$ & $A_\alpha \neq 0$\\
\hline
\hline
A.1   & $(16,0)$     & $(16,0)$     & $(16,0)$     & $(51,3)$         & 246  & $-$                & $-$ \\
\hline
A.2   & $(12,4)$     & $(8,0)$      & $(8,0)$      & $(31,7)$         & 166  & $d_{24}$           & $A_2=(S)(0^8)$, $A_4 =(V)(0^8)$ \\
\hline
A.3   & $(10,6)$     & $(4,0)$      & $(4,0)$      & $(21,9)$         & 126  & $d_{14}, d_{23}$   & $A_1=(S)(0^8)$, $A_2 =(0^8)(S)$,\\
      &              &              &              &                  &      &                    & $A_3=(0^8)(V)$, $A_4 =(V)(0^8)$ \\
\hline
A.4   & $(8,0)$      & $(8,0)$      & $(8,0)$      & $(27,3)$         & 126  & $d_{26}, d_{46}$   & $A_2 =(V')(0^8)$, $A_4 =(V)(0^8)$,\\
      &              &              &              &                  &      &                    & $A_6 =(S)(0^8)$ \\
\hline
A.5   & $(6,2)$      & $(6,2)$      & $(4,0)$      & $(19,7)$         & 106  & $d_{24}, d_{36}$   & $A_2 =(S)(0^8)$, $A_3 =(0^8)(S)$,\\
      &              &              &              &                  &      &                    & $A_4 =(V)(0^8)$, $A_6 =(0^8)(V)$\\
\hline
A.6   & $(6,2)$      & $(4,0)$      & $(4,0)$      & $(17,5)$         &  86  & $d_{16}, d_{24},$  & $A_1 =(V)(0^8)$, $A_2 =(0^8)(V)$,\\
      &              &              &              &                  &      & $d_{36}$           & $A_3 =(V')(0^8)$, $A_4 =(0^8)(S)$,\\
      &              &              &              &                  &      &                    & $A_6 =(S)(0^8)$ \\
\hline
A.7   & $(4,0)$      & $(4,0)$      & $(4,0)$      & $(15,3)$         &  66  & $d_{16}, d_{25},$  & $A_1=(V)(0^8)$, $A_2 =(0^8)(V)$,\\
      &              &              &              &                  &      & $d_{36}, d_{45}$   & $A_3 =(V')(0^8)$, $A_4 =(0^8)(V')$,\\
      &              &              &              &                  &      &                    & $A_5 =(0^8)(S)$, $A_6 =(S)(0^8)$\\
\hline
A.8   & $(3,1)$      & $(3,1)$      & $(3,1)$      & $(12,6)$         &  66  & $d_{16}, d_{24},$  & $A_1 =(W_1)(0^8)$, $A_2 =(0^8)(W_1)$,\\
      &              &              &              &                  &      & $d_{35}$           & $A_3 =(0^8)(W_1')$, $A_4 =(0^8)(W_2)$,\\
      &              &              &              &                  &      &                    & $A_5 =(0^8)(W_2')$, $A_6 =(W_2)(0^8)$\\
\hline
\end{tabular}
\end{center}
\caption{Survey of $\Z2\times\Z2$ orbifolds with generalized discrete torsion.
  The $2^\mathrm{nd}$--$4^\mathrm{th}$ columns list the number of anti-families
  and families, respectively, for the various twisted sectors $T_{(k_1,k_2)}$.
  In all models, the untwisted sector gives a contribution of $(3,3)$
  (anti-)families. $\#_S$ denotes the total number of singlets. These spectra
  can either be obtained by turning on generalized discrete torsion
  $d_{\alpha\beta}$ as specified in the next-to-last column, or by using
  lattice-valued Wilson lines $A_\alpha$ as listed in the last column. The
  building blocks are defined in equation \eqref{eq:buildingblocks}.}
\label{tab:Z2xZ2}
}}
\setlength{\captionwidth}{\oldwidth}
\end{table}

Beside other models with a net number of zero families, we find eight models
(and their mirrors, i.e.\ models where families and anti-families are exchanged).
They are listed in Tab.~\ref{tab:Z2xZ2}, where we present the number of
(anti-)families for each twisted sector and the total number of singlets. As
discussed in section~\ref{sec:GeneralizedDiscreteTorsion}, models with
non-trivial $d_{\alpha\beta}$ are equivalent to torsionless models with
lattice-valued Wilson lines. Possible representatives of these Wilson lines can
be composed out of the building blocks
\begin{align}
 W_1&  =~(0^6,1,1)\;,
 & W_2 &=~(0^5,1,1,0)\;,
 &  W_1'& =~(1,1,0^6) \;, 
 & W_2' & =~(0,1,1,0^5)\;, \nonumber \\
 S&=~(\tfrac{1}{2}^8)\;, 
 & V & =~(0^7,2)\;, 
 & V' & =~(0^6,2,0)\;,\label{eq:buildingblocks}
 & & \phantom{I^I}
\end{align}
and are listed in the last column of Tab.~\ref{tab:Z2xZ2}.

Models leading to spectra coinciding with what we got in Tab.~\ref{tab:Z2xZ2}
have already been discussed in the literature. They appeared first in
Ref.~\cite{Donagi:2004ht} in the context of free fermionic string models related
to the $\mathbbm{Z}_2\times\mathbbm{Z}_2$ orbifold with an additional freely
acting shift. More recently, new $\mathbbm{Z}_2\times\mathbbm{Z}_2$ orbifold
constructions have been found in studying orbifolds of non-factorizable
six-tori~\cite{Faraggi:2006bs,Forste:2006wq}. We find that for each model M of
Tab.~\ref{tab:Z2xZ2} there is a  corresponding `non-factorizable' model M$'$
with the following properties:
\begin{enumerate}
\item
 Each `non-empty' fixed point, i.e.\ each fixed point with local zero-modes, in the model M can be mapped
 to a fixed point with the same spectrum in model M$'$. 
\item  
 The number of `non-empty' fixed points in M coincides with the total number of
 fixed points in M$'$.
\end{enumerate}

These relations are not limited to $\mathbbm{Z}_2\times\mathbbm{Z}_2$ orbifolds,
rather we find an analogous connection also in other
$\mathbbm{Z}_N\times\mathbbm{Z}_M$ cases ($\mathbbm{Z}_N\times\mathbbm{Z}_M$
orbifolds based on non-factorizable compactification lattices have recently
been  discussed in~\cite{Takahashi:2007qc}). This result hints towards an
intriguing impact of generalized discrete torsion on the interpretation of
orbifold geometry. What the (zero-mode) spectra concerns, introducing
generalized discrete torsion (or considering generalized brother models) is
equivalent to changing the geometry of the underlying compact space,
$\Gamma\to\Gamma'$. To establish complete equivalence between these models would
require to prove that the couplings of the corresponding states are the same,
which is beyond the scope of the present study. It is, however, tempting to
speculate that non-resolvable singularities, as discussed above, do not `really'
exist as one can always choose (for a given spectrum) the compactification
lattice $\Gamma$ in such a way that there are no `empty' fixed points.

\section{How to classify $\boldsymbol{\mathbbm{Z}_{N}\times\mathbbm{Z}_{M}}$\ orbifolds}
\label{sec:ClassifZ3xZ3}

Let us now turn to describing a method of classifying heterotic
$\mathbbm{Z}_{N}\times\mathbbm{Z}_{M}$ orbifolds, taking into account
generalized discrete torsion. To illustrate our methods, we focus on
$\mathbbm{Z}_{3}\times\mathbbm{Z}_{3}$ orbifold compactifications of the
$\E8\times\E8$ heterotic string. It is straightforward to generalize the
discussion to other $\Z{N}\times\Z{M}$ orbifolds and to the \SO{32} case.


To classify an orbifold requires an efficient prescription of how to obtain all
inequivalent models. The first step in a classification is to get all admissible
choices for the shift vector $V_1$. For this purpose, we make use of Dynkin
diagram techniques (see e.g.\ \cite{Katsuki:1990bf}). These techniques are
advantageous since, when writing down $V_1$, one has the freedom of choosing the
basis of the weight lattice $\Lambda$ in such a way that the shift has a very
simple form. Clearly, this freedom is lost when one introduces the second shift
(and Wilson lines). This complicates the construction of all inequivalent shifts
$V_2$. 

To obtain all inequivalent $V_2$ we utilize a method introduced by
Giedt~\cite{Giedt:2000bi} (see also~\cite{Nilles:2006np}), i.e.\ use an adequate
minimal ansatz which avoids redundancies due to lattice translations and some
Weyl reflections. This ansatz restricts the shifts to be only in a certain cell
$\widetilde{\Lambda}_N$ of the lattice $\Lambda$ in such a way that any possible
$\mathbbm{Z}_N$ shift can be written as an element of this cell plus a lattice
vector. That is, an arbitrary $\Z{N}$ shift has a unique decomposition
\begin{equation}\label{eq:shiftdecomposition}
 V~=~\widetilde{V}+\Delta V\;,\quad \text{where}\:
 \widetilde{V}\in\widetilde{\Lambda}_N\:\text{and}\:\Delta V\in\Lambda\;.
\end{equation}
Consider now a consistent gauge embedding $(V_1,V_2)$. According to
equation~\eqref{eq:shiftdecomposition} the shifts can be decomposed into 
$(\widetilde{V}_1+\Delta V_1,\widetilde{V}_2+\Delta V_2)$ with
$\widetilde{V}_1\in\widetilde{\Lambda}_{N}$,
$\widetilde{V}_2\in\widetilde{\Lambda}_{M}$ and $\Delta V_i\in\Lambda$. It is
not hard to see that the conditions \eqref{eq:newmodularinv} imply
\begin{subequations}
\label{eq:necessaryconditions}
\begin{eqnarray}
 N\,\left(\widetilde{V}_1^2 - v_1^2\right)&~=~ 0\text{ mod }2 \;,\label{eq:necessarycondition1}\\
 M\,\left(\widetilde{V}_2^2 - v_2^2\right)&~=~0\text{ mod }2 \;,\label{eq:necessarycondition2}\\
 M\,\left(\widetilde{V}_1\cdot \widetilde{V}_2 - v_1\cdot v_2\right) &~=~0\text{ mod }1 \;.\label{eq:necessarycondition3}
\end{eqnarray}
\end{subequations}
To scan over all possible shift embeddings is therefore reduced to the task of
\begin{enumerate}
 \item specifying $\widetilde{V}_1$ (satisfying \eqref{eq:necessarycondition1})
  by Dynkin techniques, 
 \item scanning $\widetilde{\Lambda}_{M}$ for $\widetilde{V}_2$ fulfilling equations
  \eqref{eq:necessarycondition2} and \eqref{eq:necessarycondition3}, and
 \item examining all possible $(V_1,V_2)$ related to
  $(\widetilde{V}_1,\widetilde{V}_2)$ by lattice translations and satisfying
  \eqref{eq:newmodularinv}.
\end{enumerate}
At first sight, the last step seems to require a scan over an infinite number of
shifts. However, as we have seen in section~\ref{sec:DiscreteTorsion}, given one
representative $(V_1,V_2)$ satisfying \eqref{eq:newmodularinv}, this scan can be
replaced by switching on all inequivalent discrete torsion phases. Since there
is only a finite number of torsion phase assignments, we have found a
prescription to obtain all inequivalent models by scanning only over a finite set
of inputs.

As already stated, in a complete classification it is necessary to take into
account generalized discrete torsion as well. Thus, to get all admissible
shifts, one has to scan over all appropriate values for the parameters
$b_\alpha,\,c_\alpha$ and $d_{\alpha\beta}$.

All statements made for $V_2$ apply also to the Wilson lines. That is, in order
to obtain all inequivalent Wilson line embeddings, one can also scan the finite
cell $\widetilde{\Lambda}_{N_\alpha}$ (fulfilling consistency conditions
analogous to \eqref{eq:necessaryconditions}), and then switch on generalized
discrete torsion.

So far, we have described how to obtain all inequivalent models. However, some
of the inputs, specified by $(V_1,V_2)$, Wilson lines and generalized torsion
phases, turn out to be equivalent. Apart from the ambiguities related to Weyl
reflections (cf.\ \cite{Giedt:2000bi}), in the framework of $\Z{N}\times\Z{M}$
orbifolds further complications arise. For instance, we note that models with
shifts related by discrete rotations, i.e.\ 
$(V_1,V_2)\rightarrow(a_1\,V_1+a_2\,V_2,b_1\,V_1+b_2\,V_2)$ with proper values of
$a_i,b_i\in\mathbbm{Z}$, can be equivalent. For example, in  $\mathbbm{Z}_3\times\mathbbm{Z}_3$ a model
with shifts $(V_1,V_2)$ is equivalent to a model with shifts $(V_1, V_1+2\,V_2)$.

In our classification below, we consider two models as inequivalent if and only
if their massless spectra are different.\footnote{In practice, we compare the
non-Abelian massless spectra and the number of singlets. This underestimates the
true number of models somewhat.}

\subsubsection*{Sample classification of $\boldsymbol{\mathbbm{Z}_{3}\times\mathbbm{Z}_{3}}$ without Wilson lines}

As a concrete application, let us describe the classification of $\Z3\times\Z3$
orbifolds without Wilson lines.\footnote{As we allow for generalized discrete
torsion, some of the models can be interpreted as being endowed with
lattice-valued Wilson lines, see section \ref{sec:GeneralizationDTAndBrothers}.}
By using Dynkin diagram techniques, one finds that there are only five
consistent shift vectors $V_1$, which can be written in the generic form
\begin{equation}\label{eq:MinForm1}
\widetilde{V}_1~=~\frac{1}{3}(0^{n_0},1^{n_1},2^\alpha)\,(0^{m_0},1^{m_1},2^\beta),
\end{equation}
where $\alpha$ and $\beta$ can be either 0 or 1, and $n_0,n_1,m_0,m_1\in\mathbbm{Z}$, such that
$n_0+n_1+\alpha=m_0+m_1+\beta=8$. 

A general ansatz describing the second shift $V_2$ of order three is given by~\cite{Giedt:2000bi}:
\begin{eqnarray}\label{eq:MinForm2}
\widetilde{V}_2~=~ &\dfrac{1}{3}\left( \left( 
\begin{array}{c}
3\\
\vdots \\
-2
\end{array}
 \right),  \left( 
\begin{array}{c}
1\\
0
\end{array}
 \right)^{n_0-1}, \left( 
\begin{array}{c}
1\\
0\\
-1
\end{array}
 \right)^{n_1+\alpha}\;  \right)  \nonumber \\
 &\quad
\left( \left( 
\begin{array}{c}
3\\
\vdots\\
-2
\end{array}
 \right), \left( 
\begin{array}{c}
1\\
0
\end{array}
 \right)^{m_0-1}, \left( 
\begin{array}{c}
1\\
0\\
-1
\end{array}
 \right)^{m_1+\beta}\;  \right),
\end{eqnarray} 
subject to lattice conditions, equations~\eqref{eq:latticeconditions}, and to
the necessary conditions~\eqref{eq:necessaryconditions}.  Some of these models
do not fulfill the consistency requirements~\eqref{eq:newmodularinv}. As
explained above, in these cases we proceed by identifying lattice vectors
$\Delta V_i$ with the property that $(\widetilde{V}_1+\Delta
V_1,\widetilde{V}_2+\Delta V_2)$ satisfy \eqref{eq:newmodularinv}. The problem
of finding those lattice vectors can be reduced to a set of linear
Diophantine equations.

To generate all shift embeddings, we compute the spectra of models with
different values of $a$ in the discrete torsion phase. In
$\mathbbm{Z}_3\times\mathbbm{Z}_3$, the parameter $a$ can have values
$0,\,\frac{1}{3}$ or $\frac{2}{3}$. This gives a factor of three to the total
number of models. However, it turns out that not all of them are inequivalent.
Counting only inequivalent spectra, we find that there are
120 inequivalent shift embeddings.

We use now the set of shift embeddings to generate all admissible models. As
discussed in the examples of section~\ref{sec:GeneralizedDiscreteTorsion},
excluding $a$, there are 9 independent generalized discrete torsion parameters,
whose values can be again $0,\,\frac{1}{3}$ or $\frac{2}{3}$. Although the
number of models is multiplied by a factor $3^9$, the number of all inequivalent
models (spectra) is 1082. These models comprise the complete set of admissible
models without Wilson lines, or, more precisely, the complete set of models
which can be described by vanishing Wilson lines. The model definitions and the
resulting spectra are given in \cite{WebTables:2007mt}.

The above procedure can straightforwardly be carried over to the SO(32) case.
The results turn out to be similar. Repeating the steps of our $\E8\times\E8$
discussion, we find that there are 131 shift embeddings. The total number of
inequivalent models is very similar to the $\E8\times\E8$ case.

\clearpage
\section{Discussion}
\label{sec:Conclusions}

Aiming at a systematic understanding of heterotic $\Z{N}\times\Z{M}$ orbifolds,
we have investigated the possibilities arising in these constructions. We find
that, unlike in the case of prime orbifolds, adding elements of the weight
lattice to the shifts, $(V_1,V_2)\to(V_1+\Delta V_1,V_2+\Delta V_2)$, changes in
general the spectrum. Interestingly, the same spectra are obtained by switching
on discrete torsion. Stated differently, one can trade discrete torsion for a
change of the gauge embedding by lattice vectors.   
We have extended our analysis such as to include generalized discrete torsion.
We find that a large class of generalized discrete torsion phases can be
mimicked by lattice-valued Wilson lines. Interestingly, an analog of a
generalized discrete torsion phase appears in certain \Z{N} orbifolds which
admit at least two discrete Wilson lines of coinciding orders. Another
remarkable result is that switching on certain types of generalized discrete
torsion is equivalent to changing the 6D compactification lattice.  This implies
that the field-theoretic interpretation of the model input can be somewhat
subtle. We provided various explicit examples, illustrating all main statements.

Our results have important consequences. At the more practical side, we were
able to formulate a straightforward method of classifying $\Z{N}\times\Z{M}$
orbifolds. We have also seen that switching on generalized discrete torsion
allows to change the local spectra, i.e.\ one obtains different twisted states,
which correspond to brane fields in the field-theoretic description. We expect
this to become important for orbifold model building, where one can use this
observation, for instance, for reducing the number of generations without
modifying the gauge group.

At a more conceptual level, our findings imply that a given spectrum cannot be
ascribed neither to properties of the 6D internal space alone, i.e.\ whether
discrete torsion is switched on or not, nor to the gauge embedding, but only to
both. This implies that the same models, leading to the same spectra, can be
regarded as resulting from what one might consider as different geometries.
Although we cannot claim to have identified the deeper reasons for these
relations, we feel that our observations constitute some progress in the task of
understanding stringy geometry.

\paragraph{Acknowledgments.}
We would like to thank S.~F\"{o}rste, A.~Micu and H.~P.~Nilles for valuable
discussions, and T.~Kobayashi and K.-J.~Takahashi for correspondence.

This work was partially supported by the cluster of  excellence ``Origin and
Structure of the Universe'', the EU 6th Framework Program MRTN-CT-2004-503369 ``Quest
for Unification'', MRTN-CT-2004-005104 ``ForcesUniverse'',  MRTN-CT-2006-035863
``UniverseNet'' and the Transregios 27 '`Neutrinos and Beyond'' and  33 ``The Dark
Universe'' by Deutsche Forschungsgemeinschaft (DFG).

\appendix
\renewcommand{\theequation}{\Alph{section}.\arabic{equation}}
\setcounter{section}{0}
\setcounter{equation}{0}
\renewcommand{\thetable}{\Alph{section}.\arabic{table}}
\clearpage

\section{Transformation phases}
\label{app:ModularInvariance}

The aim of this appendix is to clarify the transformation law of physical
states. Let us start with the simplest example, a $\Z{N}$ orbifold without
Wilson lines. Modular invariance requires \cite{Vafa:1986wx}
\begin{equation}\label{eq:weakmodularinv}
 N\,(V^2-v^2)~=~0\mod 2\;.
\end{equation}
To see that there are some subtleties,  consider a `constructing element'
$g=(\theta^k,n_\alpha e_\alpha)$. Zero-modes in the $g$-twisted sector have to
fulfill the masslessness condition \eqref{eq:MassEquation}. Focus, for
simplicity, on non-oscillator states, for which the masslessness conditions read
\begin{subequations}
\begin{eqnarray}
 \frac{1}{2}p_\mathrm{sh}^2-1+\delta c & = & 0\;,\\
 \frac{1}{2}q_\mathrm{sh}^2-\frac{1}{2}+\delta c & = & 0\;,
\end{eqnarray}
\end{subequations}
where $p_\mathrm{sh}=(p+V_g)$, $q_\mathrm{sh}=(q+v_g)$, $V_g=k\,
V+n_\alpha\,A_\alpha$ and $v_g=k\,v$. Using the properties $p^2=\text{even}$ and
$q^2=\text{odd}$ one derives
\begin{subequations}\label{eq:p.V}
\begin{eqnarray}
 p\cdot V_g & = & 1-\delta c-\frac{V_g^2}{2}+\text{integer}\;,\\
 q\cdot v_g & = & -\delta c-\frac{v_g^2}{2}+\text{integer}\;.
\end{eqnarray}
\end{subequations}
Now let us study the transformation of a solution of the mass equation
$|\psi\rangle$. The shifted momenta $p_\mathrm{sh}$ and $q_\mathrm{sh}$ represent the gauge
and Lorentz (or $\SO8$) quantum numbers. This fixes the transformation phase of
the $g$-twisted string $|\psi\rangle$ to be
\begin{equation}\label{eq:ProjectionCondition2}
 2\pi\,[p_\mathrm{sh}\cdot V_g-q_\mathrm{sh}\cdot v_g]~=~
 2\pi\,\left[\frac{1}{2}\left(V_g^2-v_g^2\right)\mod 1\right]
\end{equation}
under the action of $g$. From \eqref{eq:weakmodularinv} we infer that this phase
does not vanish in general, rather it is of the form $2\pi\,m/N$ with
$m\in\mathbbm{Z}$. This raises the question of how a state associated to $g$ can
be invariant under the action of $g$. In the literature, the transformation
behavior of states associated to twisted strings has often been `repaired',
i.e.\ an additional transformation phase has been introduced by hand. In what
follows, we present a geometric explanation of how such additional phases arise.

\begin{figure}[!h]
\centerline{\CenterObject{\includegraphics{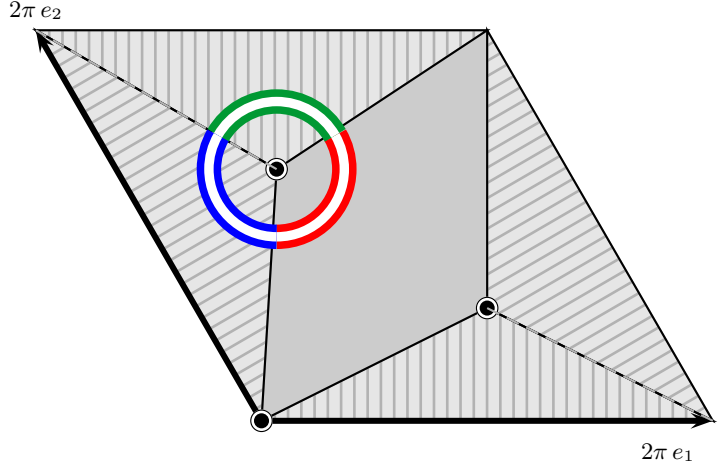}}}
\caption{$\mathbbm{T}^2_{\SU3}/\Z3$ orbifold. The fundamental domain of the
torus can be taken to be the darker area. Transformation of
$g=(\theta,e_1+e_2)$-twisted strings under (the constructing element) $g$: the
(red) right string is mapped to the (green) upper, and the (green) upper to the
(blue) left, and the (blue) left back to the (red) upper one.}
\label{fig:Transformation}
\end{figure}

Consider a string twisted by the space group element $g$ in the `upstairs'
picture, i.e.\ on the torus. Twisted strings end at the borders of the
fundamental domain of the orbifold. The fundamental domain of the orbifold
together with its non-trivial images under $g$, $g^2$ etc.\ comprise the
fundamental domain of the torus. Therefore, on the torus a twisted string
appears in $n$ copies where $n$ is the order of $g$, i.e.\ the minimal positive
integer with $g^n=\mathbbm{1}$. A 2D illustration is shown in
figure~\ref{fig:Transformation}.
That is, a $g$-twisted state appears as a linear combination of the state
$|\psi\rangle$ corresponding to the $g$-twisted string in the fundamental domain
and its images under $g^m$. This linear combination can involve phase factors
with the constraint being that $g^n$ acts as identity. There are $n$
different linear combinations labeled by $m\in\mathbbm{Z}$,
\begin{equation}\label{eq:LinComb}
 |\psi^{(m)}\rangle~=~
 \frac{1}{\sqrt{n}}
 \left(|\psi\rangle+e^{-2\pi\I\, m/n}\,|g\,\psi\rangle
 + \dots + e^{-2\pi\I\, m\,(n-1)/n}\,|g^{n-1}\,\psi\rangle\right)\;,
\end{equation}
on the torus. In figure~\ref{fig:Transformation} this would correspond to a
superposition of the red, green and blue string, weighted by phase factors.  It
is clear that, under the action of $g$, such a linear combination picks  up a
phase $e^{2\pi\I\, m/n}$.  In other words, the transformation phase
\eqref{eq:ProjectionCondition2} is to be amended by $2\pi\, m/n$.

It is straightforward to apply these observations to the above-mentioned problem
of $g$-invariance of states associated to the constructing element $g$. Clearly,
only the linear combination \eqref{eq:LinComb} with $m=-\frac{n}{2}(V_g^2-v_g^2)$
is $g$-invariant. So we conclude that for every solution of the mass equation
one finds precisely one $g$-invariant state. In addition to the phase arising
from the gauge and Lorentz quantum numbers \eqref{eq:ProjectionCondition2}, this
state picks up a compensating phase 
\begin{equation}\label{eq:PhiVac}
 \Phi_\mathrm{vac}~=~ \exp\left\{2\pi\I\,\left[-\frac{1}{2}(V_g\cdot V_g-v_g\cdot v_g)\right]\right\}
\end{equation}
under the action of $g$. 

Before turning to the discussion of the transformation of such states under
commuting elements $h$, let us briefly comment on a technical simplification
that is possible in many models.
It is also clear that, in \Z{N} orbifolds without Wilson lines, one can
`transform' a model M, with a `weak' shift $V$, to the model M$'$, with shift
$V'= V+\Delta V$ where $\Delta V\in\Lambda$. The solutions of the mass equation
coincide in the models M and M$'$. It is, up to some exceptional cases which we
will discuss below, always possible to find a $\Delta V$ with
$\frac{1}{2}[(V+\Delta V)^2-v^2]\in\mathbbm{Z}$ so that
\eqref{eq:ProjectionCondition2} is automatically fulfilled for any
$|\psi\rangle$ (and therefore also for trivial linear combinations). 
For a large class of constructions one can therefore adopt the following logic.
Models with an input fulfilling only the (`weak') modular invariance constraints
\eqref{eq:weakmodularinv} might not be considered as they have an alternative
description in terms of a model with input fulfilling the stronger constraints
\begin{equation}\label{eq:strongmodularinv}
V^2-v^2~=~0\mod 2\;.
\end{equation}
That is, in order to avoid double-counting, one can restrict to the stronger
constraints \eqref{eq:strongmodularinv} in many cases. Similar statements
apply to the case with non-trivial Wilson lines (an example has been given in
\cite{Buchmuller:2006ik}).
However, there is a caveat, namely the above-mentioned exceptional cases in
which a `weak' modular invariant input cannot be transformed to the `strong'
form. The simplest example for such a case is a \Z3 orbifold with $V=0$. Further
examples arise in non-prime \Z{N\cdot M} orbifolds ($N,M>1$) where $N\cdot
V\in\Lambda$. 

Let us now return to the question of how a ($g$-invariant) state associated to
the constructing element $g$ transforms under the action of a commuting element
$h$. In the following, we denote the corresponding transformation phase of
equation~\eqref{eq:transformationphase} by $\Phi(g,h)$. Clearly, this phase has to comply with the 
space group multiplication law, thus
\begin{equation}
  \Phi_\mathrm{vac}(g,\,g^p\,h^q)
  ~=~
  [\Phi_\mathrm{vac}(g,\,g)]^p\,[\Phi_\mathrm{vac}(g,\,h)]^q\;.
\end{equation}
Since $\Phi_\mathrm{vac}(g,\,g)$ is already fixed by \eqref{eq:PhiVac}, this leads
to the conclusion
\begin{equation}
 \Phi_\mathrm{vac}(g,\,h)
 ~=~
 \exp\left\{2\pi\I\,\left[-\frac{1}{2}\left(V_g\cdot V_h-v_g\cdot v_h\right)\right]\right\}
\end{equation}
for any commuting $h\in S$. Therefore, the full transformation phase of the physical states has to be defined as in
equation~\eqref{eq:transformationphase}. But there are still some constraints which have to be fulfilled
for the sake of consistency. To illustrate them, let us consider $g,h \in S$ with $g^n = \mathbbm{1} =
h^s$. From the definition of the full transformation phase $\Phi$, it is clear that one has to demand
\begin{equation} \label{eq:TransfModified}
\Phi(g,h) ~\stackrel{!}{=}~ \Phi(g^{n+1},h)
~=~\Phi(g,h)\,\Phi_\mathrm{vac}(g,h)^{-n} \,,
\end{equation}
where the second equality is obtained by replacing, according to the
usual embedding, $V_g\to (n+1)\,V_g$ and $v_g\to(n+1)\,v_g$ in equation~\eqref{eq:transformationphase}.
Thus, $\Phi_\mathrm{vac}(g,h)^{n}\stackrel{!}{=}1$. An analogous
reasoning starting with $\Phi(g,h^{s+1})$ leads to
$\Phi_\mathrm{vac}(g,h)^{s}~\stackrel{!}{=}~1$ and thus finally to 
\begin{equation}\label{eq:consistency}
\Phi_\mathrm{vac}(g,h)^{\mathrm{gcd}(n,s)}~\stackrel{!}{=}~1\;.
\end{equation}
Formulating equation~\eqref{eq:consistency} in terms of the gauge embedding
shifts leads to the consistency conditions~\eqref{eq:newmodularinv} on shifts
and Wilson lines.\footnote{In the case of two different \Z2 Wilson lines we find
that \eqref{eq:fsmi5} can be relaxed, i.e.\ $\text{gcd}(N_{\alpha},N_{\beta})$
can be replaced by $N_{\alpha}\,N_{\beta}=4$, provided there exists no  $g \in
P$ with the property $g\,e_\alpha~\neq~e_\alpha$ but $g\,e_\beta~=~e_\beta$.
Imposing the weaker condition leads, as we find, to anomaly-free spectra.}

\providecommand{\bysame}{\leavevmode\hbox to3em{\hrulefill}\thinspace}

\end{document}